\documentclass[10pt]{article}
\usepackage{amssymb,epsfig}
\usepackage{times}
\newcommand{\Ga}{\Gamma}
\newcommand{\Si}{\Sigma}
\newcommand{\de}{\delta}
\newcommand{\bP}{{\mathbb P}}
\newcommand{\bQ}{{\mathbb Q}}
\newcommand{\psq}{{\bar\psi}}

\newcommand{\veps}{\varepsilon}
\newcommand{\ps}{\psi}
\newcommand{\Ref}[1]{$(\ref{#1})$}
\newcommand{\sfrac}[2]{{\textstyle \frac{#1}{#2}}}
\newcommand{\cL}{{\cal L}}
\newcommand{\cA}{{\cal A}}
\newcommand{\cD}{{\cal D}}
\newcommand{\del}{\partial}
\newcommand{\al}{\alpha}
\newcommand{\si}{\sigma}
\newcommand{\om}{\omega}
\newcommand{\De}{\Delta}
\newcommand{\be}{\beta}
\renewcommand{\th}{\theta}
\newcommand{\ga}{\gamma}
\newcommand{\cE}{{\cal E}}
\newcommand{\bU}{{\mathbb U}}
\newcommand{\bA}{{\mathbb A}}

\newcommand{\gtoas}[1]{{\quad\mathop{\longrightarrow}\limits_{#1}\quad}}
\newcommand{\et}{\eta}
\newcommand{\etq}{{\bar\eta}}
\newcommand{\ph}{\phi}
\newcommand{\phq}{{\bar\phi}}
\newcommand{\vphi}{\varphi}
\newcommand{\Om}{\Omega}
\newcommand{\cR}{{\cal R}}
\newcommand{\cT}{{\cal T}}

\newcommand{\E}{{\rm e}}
\newcommand{\I}{{\rm i}}  
\newcommand{\dd}{{\rm d}}
\newcommand{\Esc}[1]{\epsilon_{#1}}
\newcommand{\Pauli}{\sigma}

\renewcommand{\vec}[1]{{\bf #1}}

\newcommand{\Tr}{\mathop{\mbox{Tr}}}
\newcommand{\Four}{V}
\newcommand{\Ebrackl}{{\lbrack\!\lbrack}}
\newcommand{\Ebrackr}{{\rbrack\!\rbrack}}
\newcommand{\Equal}[1]{\Ebrackl #1 \Ebrackr}
\newcommand{\BEqual}[1]{{\Big\lbrack}\!\!{\Big\lbrack} #1 {\Big\rbrack}\!\!{\Big\rbrack}}
\newcommand{\X}{X}
\newcommand{\Xq}{\bar X}
\newcommand{\Lap}{\triangle}
\newcommand{\bigQ}{\bQ}
\newcommand{\lQ}{{\rm q}}
\newcommand{\sQ}{Q}
\newcommand{\SY}{\bU}
\newcommand{\SG}{\bA}
\newcommand{\bDe}{{\bf \De}} 
\newcommand{\ovl}[1]{\overline{#1}}

\begin{document}
\title{Renormalization group flows into phases with broken symmetry}

\author{Manfred Salmhofer$^1$, Carsten Honerkamp$^2$, Walter Metzner $^2$,
\\
Oliver Lauscher $^1$ 
\\ 
\small 
$^1$ Institut f\" ur Theoretische Physik, Universit\"{a}t Leipzig, 
\\ \small
Augustusplatz 10, D-04109 Leipzig, Germany \\
\small
$^2$ Max--Planck--Institut f\" ur Festk\" orperforschung, 
\\ \small
Heisenbergstrasse 1, D-70569 Stuttgart, Germany
}

\date{\today}

\maketitle
\normalsize 
\begin{abstract}
\noindent
We describe a way to continue the fermionic renormalization group flow 
into phases with broken global symmetry. The method does not require 
a Hubbard-Stratonovich decoupling of the interaction.
Instead an infinitesimally small symme\-try-breaking component 
is inserted in the initial action, as an initial condition 
for the flow of the selfenergy. Its flow is driven by the interaction 
and at low scales it saturates at a nonzero value
if there is a tendency for spontaneous symmetry breaking 
in the corresponding channel. 
For the reduced BCS model we show how a small 
initial gap amplitude flows to the value 
given by the exact solution of the model. 
We also discuss the emergence of the Goldstone boson in this approach.
\end{abstract}

\section{Introduction} \label{sec:intro}
Renormalization group (RG) flows for interacting fermions have experienced a massive upswing in recent years
\cite{ZS,HM,HSFR,SalHon,Tflow,And,Bourbonnais,KatKam,Tsai1,Schwenk,HonSalZ}. 
Besides their conceptual importance for the understanding of many-particle systems they have become a successful unbiased vehicle for the detection and evaluation of Fermi surface instabilities in low-dimensional many-fermion systems. For example, the two-dimensional Hubbard model has been subject to numerous RG studies over the last years. For most parameter regions in the weak coupling range of this model, it has led to a good qualitative picture of the dominant instabilities of the Fermi liquid state. In most cases the  phases suggested by the RG are states with spontaneously broken symmetry, like superconducting or magnetically ordered states.
Yet a major drawback of the existing weak--coupling fermionic RG schemes 
is their failure to give a controlled access to the symmetry-broken regime.  
In the language of the fermionic RG used up to now,
the instabilities towards these states give rise to a flow to strong coupling at some small energy scale where at least one class of coupling constants grows larger than the bandwidth of the system. Then the weak--coupling RG becomes
unreliable, and if one ignores this and tries to continue the flow, 
the coupling functions diverge before all degrees of freedom 
have been integrated over.
Hence, so far the only way to develop a theory for 
the strongly coupled phase is to stop the fermionic 
RG flow before it becomes unreliable
and to resort to some other method for the scales below, 
e.g., a mean-field treatment, bosonization \cite{Lin,Solyom,Delft}, 
or exact diagonalization of a heavily reduced model \cite{Lauchli}.
The antiferromagnetic phase of the half--filled Hubbard model 
was studied with bosonic RG methods in \cite{Baier}.

Here we propose an extension of the fermionic functional RG methods which allows us to access the symmetry-broken regime continuously. 
The basic idea is to insert a small symmetry-breaking component into the initial condition of the flow at the initial energy scale $\Esc{0}$.
In a magnetic system, this would be a small external field, 
and if spontaneous symmetry breaking happens, 
the spontaneous magnetization is obtained in the limit
where this field vanishes. For the case of a Cooper instability towards a superconducting state the symmetry-breaking term is  
a small superconducting gap amplitude $\Delta_0$. 
In the RG flow, degrees of freedom with energy above a scale 
$\Esc{s}=\Esc{0} \E^{-s}$ are gradually integrated out,
and the full model is recovered in the limit $s \to \infty$, 
where $\Esc{s} \to 0$ ($\Esc{0}$ is a fixed energy scale; 
we take it equal to the bandwidth of the model). 
Now in the course of the flow, the gap amplitude gets renormalized 
by the flow equation for the self-energy, so that 
$\Delta_s$ tends to a ($\Delta_0$--dependent) final value
$\Delta_{\rm final}$ in the limit of vanishing energy scale, $s \to \infty$. 
We will show below that, together with a reorganization of the hierarchy 
of flow equations for the one-particle irreducible vertex function 
proposed recently by Katanin \cite{Katanin}, 
this scheme allows us to reproduce the {\em exact} mean-field results 
for the reduced BCS model. 
That is, we can take the limit $s \to \infty$, hence
integrate over all degrees of freedom without encountering 
a divergence in the coupling functions, and when
the external field $\Delta_0$ is sent to zero after that,
the final gap $\Delta_{\rm final}$ 
converges to the solution of the BCS equation. Moreover, the 
two--particle vertex also converges to the exact result for the BCS model.
Furthermore we show that the Ward identity from global $U(1)$ charge symmetry
is satisfied in our approach. This Ward identity, together with the 
solution for the gap, fixes the two--particle vertex of the model,
and we recover the divergence at zero energy which is the remnant
of the Goldstone mode in the reduced BCS model.  

In this paper we restrict our detailed calculations to the reduced BCS model, 
but we emphasize that our method is also applicable and expected 
to give reasonable results for models where no exact results are available 
and where no mean-field theory can be developed straightforwardly. 
This will be discussed in Sec. \ref{sec:genmod}. 

The structure of the paper is follows. In Sec. \ref{sec:rgform} we briefly go through the essential points in the RG formalism for the one-particle irreducible (1PI) vertex functions the further treatment is based on. We describe a modification of the scheme proposed recently by Katanin\cite{Katanin} which allows one to reproduce the selfconsistent random phase approximation (RPA) + Hartree resummation in a certain channel. In Secs. \ref{sec:sc} and \ref{sec:gf} we apply this modified scheme to the reduced BCS model and show how it allows us to flow into the symmetry-broken regime. We also show how the Ward identity is recovered for this case. The outlook in Sec. \ref{sec:genmod} briefly discusses the superconducting gap flow in models more general than the reduced BCS model. The appendix contains some details on the derivation of the gap flow in Sec. \ref{sec:gf}.

\section{The renormalization group flow} \label{sec:rgform}
\subsection{The flow for the 1PI functions}\label{sec:1pi}
We briefly recall the flow equations for the 1PI vertices of a general 
fermionic  model, as described in \cite{SalHon} (for setup and notations,
see \cite{SalHon}). 
The generating functional $W(H)$ of the connected Green functions 
generated by sources $H$ is defined as 
\begin{equation}\label{eq:Wdef}
\E^{-W(Q,H)} 
=
\int \cD \Psi \;
\E^{-\frac12(\Psi,Q\Psi) - V_0(\Psi) + (H,\Psi)} .
\end{equation}
where the quadratic part of the action, $Q$, defines the free propagator
and we assume that the interaction $V_0$ is short range and even,
$V_0(-\Psi) = V_0(\Psi)$. We have explicitly denoted the dependence 
of $W$ on $Q$ here. Most of the time we shall abbreviate $W(Q,H) = W(H)$;
the $Q$--dependence will resurface in the notation when we discuss
the Ward identities. 

In an RG flow, $Q=Q_s$ depends on a scale parameter $s$. 
Hence $W$ also depends on $s$ and 
it satisfies the differential equation
\footnote{The form stated in (37) of \cite{SalHon}
follows by straightforward differentiation,
(up to an inessential field--independent term arising from
the different normalization of the Grassmann Gaussian 
integral in the definition of $W$). }
(see \cite{SalHon})
\begin{equation}\label{eq:WPolch}
\dot W (H)
=
-
\E^{W}
\frac12
\left(
\frac{\de}{\de H}\, , \; \dot Q \;
\frac{\de}{\de H}
\right) \E^{-W}
\end{equation}
Here the dot denotes the derivative with respect to $s$. 
The Legendre transform $\Ga$ of $W$ also depends on $s$,
and it satisfies the differential equation (see \cite{SalHon})
\begin{equation}\label{eq:RGDE1}
\dot \Ga (\Psi) 
=
\frac12 (\Psi, \dot Q \Psi) 
+
\frac12 \Tr \left[ 
\dot Q \left(\frac{\de^2 \Ga}{\de \Psi^2}\right)^{-1} 
\right] .
\end{equation}
This nonpolynomial equation can be transformed into a 
hierarchy of differential equations in which the vertex functions 
appear only polynomially by expanding in the fields:
\begin{equation}
\Ga(\Psi) 
=
K + 
\frac12 (\Psi,\tilde\Ga^{(2)} \Psi) + \Ga ^{(\ge 4)} (\Psi)
\end{equation}
where 
\begin{equation}
\Ga ^{(\ge 4)} (\Psi)
=
\sum_{m\ge 2}
\Ga^{(2m)} (\Psi)
\end{equation}
and $\Ga^{(2m)}$ is homogeneous of degree $2m$ in $\Psi$.
There are no odd terms in this expansion because $V_0$ is even in $\Psi$. 
As in \cite{SalHon} we denote the derivative appearing  
in \Ref{eq:RGDE1} by
\begin{equation}
\tilde \Ga_{XX'} 
=
\frac{\de^2 \Ga}{\de\Psi_X\de \Psi_{X'}}
=
-\tilde \Ga^{(2)}_{XX'}
+
\tilde \Ga^{(\ge 4)}_{XX'}
\end{equation}
For $V_0=0$, $\tilde\Ga^{(2)} = Q$. For $V_0 \ne 0$ 
we define the selfenergy $\Si$ by 
\begin{equation}
\tilde \Ga^{(2)} = Q - \Si .
\end{equation}
The full propagator is 
\begin{equation}\label{eq:G}
G = \left( \tilde\Ga^{(2)}\right) ^{-1}  = 
(Q - \Si)^{-1} .
\end{equation}
With this, 
\begin{equation}
\tilde \Ga^{-1} 
=
- \left( 1 - G \tilde \Ga^{(\ge 4)} \right)^{-1} \, G .
\end{equation}
Comparing homogeneous parts in $\Psi$ we arrive at the set of equations
\begin{eqnarray}
\dot K 
&=&
-\frac12 \Tr (\dot Q G)
\nonumber
\\
\dot \Si 
&=&
- \frac12 
\Tr (S \tilde \Ga^{(\ge 4)})
\\
\dot \Ga^{(2m)}
&=&
\frac12 \Tr (S \tilde\Ga^{(2m+2)})
+
\frac12
\bP_{2m}
\Tr \left(S\tilde \Ga^{(\ge 4)}G\tilde \Ga^{(\ge 4)}
(1-G\tilde \Ga^{(\ge 4)})^{-1}\right)
\nonumber
\end{eqnarray}
where 
\begin{equation}
S = - G \dot Q G
\end{equation}
is the single scale propagator and 
$\bP_{2m}$ denotes the projection on the degree $2m$ part 
of any element of the Grassmann algebra. 
Because every factor $\tilde \Ga^{(\ge 4)}$ increases the power of $\Psi$
by $2$, only finitely many terms appear on the right hand side of the 
equation for $\dot \Ga^{(2m)}$.
The equation for $\Four = \Ga^{(4)}$ reads
\begin{equation}\label{eq:Ga4DE}
\dot \Four
=
\frac12 \Tr (S \tilde\Ga^{(6)})
+
\frac12
\Tr \left(S\tilde \Four G\tilde \Four \right) .
\end{equation}
For $m=3$ we have
\begin{eqnarray}\label{eq:Ga6DE}
\dot \Ga^{(6)}
&=&
\frac12 \Tr (S \tilde\Ga^{(8)})
\\
&+&
\frac12
\left(
\Tr (S\tilde \Ga^{(6)}G\tilde \Four)
+
\Tr (S\tilde \Four G\tilde \Ga^{(6)})
+
\Tr (S\tilde \Four G\tilde \Four G\tilde \Four)
\right) .
\nonumber
\end{eqnarray}
In general, the flow equations have the property that on
the right hand side of the equation for $\dot \Four$, 
a term quadratic in $\Four$ appears, but that
for all $m \ge 3$, $\Ga^{(2m)}$ appears at most linearly 
on the right hand side of the equation for $\dot \Ga^{(2m)}$.
The term linear in $\Ga^{(2m)}$ leads to a resummation
of four--point insertions which includes ladder summations,
as described in \cite{SalHon}.

It is thus natural to count powers in an expansion 
in the renormalized coupling function $\Four$.
We assume that the initial interaction at $s=0$
consists of only a selfenergy term and a four--point vertex 
$\Four_0$. By forming a tree and adding one extra loop line
we see that a 1PI graph with $r$ four--legged vertices 
can have at most $4r-2(r-1) -2 = 2r$ external legs. 
Because $\Ga^{(2m)}$ is a sum of 1PI graphs with
$2m$ external legs, it is at least of order $m$ in $\Four$. 
Thus, to order $m$ in $\Four$, we can drop $\Ga^{(2m+2)}$. 
Then the equation for $\Ga^{(2m)}$ closes. 
This procedure is described in detail for the second and
third order flow in Section \ref{se:3rdord}; 
it generalizes straightforwardly to arbitrary orders. 

For an antisymmetric operator $A=-A^T$, define the functional Laplacian 
by 
\begin{equation}\label{eq:Lapdef}
\Lap_A = 
\left(
\frac{\de}{\de \Psi}, \; 
A \frac{\de}{\de \Psi}
\right)
=
\int \dd X \int \dd Y 
\frac{\de}{\de \Psi(X)} \; 
A (X,Y) \frac{\de}{\de \Psi(Y)} .
\end{equation}
Laplacians can be used to write the traces in a convenient form;
we shall in the following use that
\begin{equation}\label{eq:TrLap}
\Tr (S \tilde\Ga^{(2m+2)})
=
-\Lap_S \Ga^{(2m+2)} .
\end{equation}
The minus sign appears because $S$ is antisymmetric. 
Representations by Laplacians can also be used to rewrite the
other terms in the flow equation, by introducing copies of the fields.
This will be useful in the derivations of the next section,
so we discuss it here.  
We introduce two (or more) copies $\Psi^{(1)}, \Psi^{(2)}, \ldots $
of the field $\Psi$. These copies serve to simplify applications of the 
product rule and the distinction of which derivatives act on which function. 
Since the copies are only a combinatorial device, we shall in the end
evaluate at $\Psi^{(1)} = \Psi^{(2)}= \ldots = \Psi$. 
We denote this operation by $\Equal{ \;\cdot\;}$, e.g.\
\begin{equation}
\BEqual{A(\Psi^{(1)}) B(\Psi^{(2)}) C(\Psi^{(3)}) }
= A(\Psi) B(\Psi) C(\Psi) .
\end{equation}
With this and the notation
\begin{equation}
\Lap_C^{(i,j)} 
= 
\left( \frac{\de}{\de\Psi^{(i)}}, \; C \frac{\de}{\de\Psi^{(j)}}\right)
\end{equation}
we can write e.g.\
\begin{equation}\label{eq:Lapcop}
\Tr (S \tilde V G \tilde V) 
=
- \BEqual{\Lap_S^{(1,2)} \Lap_G^{(1,2)} V^{(1)} V^{(2)} } .
\end{equation}
Here we also used the abbreviation $V^{(i)} = V(\Psi^{(i)})$.

\subsection{The second and third order ${\bf \dot G}$ scheme}\label{se:3rdord}
In this section, we justify the modification  proposed in \cite{Katanin},
namely putting $\dot G$ propagators instead of single scale propagators
$S$ on the lines in the 1PI hierarchy,
%(except in the equation for $\dot\Sigma$), 
and describe a systematic procedure for doing this. 
In second order, this replacement is simply
\begin{equation}\label{eq:2Four}
\dot \Four = 
- \frac12 \BEqual{\Lap_G^{(1,2)} \Lap_{\dot G}^{(1,2)} 
\Four^{(1)} \Four^{(2)} }
=
- \frac14 \BEqual{\left(\sfrac{\del}{\del s} {(\Lap_G^{(1,2)})}^2\right)
\Four^{(1)} \Four^{(2)} } .
\end{equation}
We show below that this suffices to do flows with symmetry breaking 
that recover the BCS solution exactly. 
However, a further discussion is necessary because
this replacement is to some extent ad hoc. 
Indeed,  we shall see that
its systematic justification requires going 
to all orders in the expansion in the renormalized coupling. 
To any fixed order in $V$, the replacement is not unambiguous
because the difference $\dot G- S  = G\dot \Si G$ is at least of first 
order in $V$. Thus, in any contribution of order $m$, a 
single--scale propagator $S$ associated to a line can be replaced
by a differentiated full propagator $\dot G$ up to orders $V^{m+1}$ 
-- or one may as well leave it as 
a single--scale propagator $S$ up to orders $V^{m+1}$. 
The replacement of $S$ by $\dot G$ may be more suited for 
self--consistency arguments, 
but in a treatment where powers of $V$ are kept only up to a fixed
order $m$, one cannot chose one over the other. 
The ambiguity is removed if one realizes that one can 
rewrite the 1PI hierarchy to all orders 
in a way where only full four-point functions $V$ appear as vertices, 
and full propagators $G$ and
differentiated full propagators $\dot G$ appear on the lines,
and where all tadpole terms are removed,
except in the equation for $\dot \Si$ itself 
(i.e.\ the lowest equation in the hierarchy).
We give the explicit equations to third order in $V$.
They have a simple and nice form which will be useful 
for doing flows which fully take into account the 
vertex corrections that one needs to do in a full 
two--loop calculation of  the selfenergy flow, 
which has not been done up to now. 

As already discussed above, we assume in the following
that the initial conditions
for the flow are that $\Ga^{(\ge 6)} = 0$ at $s=0$, and that 
$G(0) =0$ because the cutoff function is such that $s=0$ corresponds 
to energies outside the band range. 

The derivative of \Ref{eq:G} gives  
\begin{equation}\label{eq:dotG}
\dot G 
=
S + G \dot \Sigma G .
\end{equation}
The initial condition for $\Ga^{(6)}$ is zero, so 
\Ref{eq:Ga6DE} implies that $\Ga^{(6)}$ is at least of third order
in $\Four$. The terms in which $\Ga^{(6)}$ and $\Four$
appear in \Ref{eq:Ga6DE} are therefore at least of fourth order,
and to third order, the equation for $\Ga^{(6)}$ simplifies to
\begin{equation}
\dot \Ga^{(6)}
=
\frac12
\Tr (S\tilde \Four G\tilde \Four G\tilde \Four) .
\end{equation}
By \Ref{eq:dotG}, replacing $S$ by $\dot G$ in this equation 
does not change anything to third order since $\dot \Si$ is at least of first order
in $\Four$. 
By \Ref{eq:Ga4DE}, $\dot \Four$ is at least of second order in $\Four$,
so 
\begin{equation}
\dot \Ga^{(6)}
=
\frac16
\frac{\del}{\del s}
\Tr (G\tilde \Four G\tilde \Four G\tilde \Four) .
+
O\left(\Four^4\right)
\end{equation}
Discarding the fourth and higher order terms we can integrate this equation. 
Because $G(0)=0$  and $\Ga^{(6)} (0) = 0$, 
\begin{equation}\label{eq:6pf}
\Ga^{(6)}
=
\frac16
\Tr (G\tilde\Four G\tilde\Four G\tilde\Four) .
\end{equation}
Inserting this into \Ref{eq:Ga4DE} and using \Ref{eq:TrLap}, we get
\begin{equation}\label{eq:Fournoamoi}
\dot \Four
=
\frac12
\left[
\Tr \left(S\tilde \Four G\tilde \Four\right)
-
\frac{1}{2}
\Lap_{S}
\frac{1}{3}
\Tr (G\tilde\Four G\tilde\Four G\tilde\Four) 
\right] .
\end{equation}
If both derivatives of $\frac12\Lap_S$ act on the same factor $\tilde\Four$,
the result is $\frac12\Lap_S \tilde\Four$ which equals $\dot \Si$.
There are three such terms, and they all give the same contribution
by cyclicity of the trace. Thus the $1/3$ gets canceled and by \Ref{eq:dotG}, we can combine 
this term with the second order term in \Ref{eq:Fournoamoi},
to replace $S$ by $\dot G$.  
Using \Ref{eq:Lapcop}, 
we obtain the flow equations that are exact to third order in $\Four$ as
\begin{eqnarray}
\dot \Four
&=&
- \frac14
\BEqual{\left( \frac{\del}{\del s} (\Lap_G^{(1,2)})^2\right)  \Four^{(1)} \Four^{(2)} } 
\label{eq:3Four}
\\
&&
-
\frac14
\frac{\del}{\del s}
\BEqual{\Lap_G^{(1,2)}\Lap_G^{(1,3)}(\Lap_G^{(2,3)})^2\;
\Four^{(1)} \Four^{(2)}\Four^{(3)}  }
\nonumber
\\
\dot \Si 
&=& 
 \frac12 \Lap_S \tilde \Four .
\label{eq:3Si}
\end{eqnarray}
The change in third order, as compared to \Ref{eq:2Four}, is 
simply the $s$--derivative of a single third order diagram. 
One can also rewrite the equation for the selfenergy such that
no $S$--propagator appears any more:
by \Ref{eq:dotG},  \Ref{eq:3Si} becomes
\begin{equation}
(\Psi, \dot \Si \Psi) + \Lap_{ G \dot \Si G} \Four
=
- \frac12 \Lap_{\dot G} \Four
\end{equation}
If we can solve these equations, the 1PI six-point function is given by 
\Ref{eq:6pf} and $\Ga^{(2m)}= 0$ for all $m \ge 4$. That is, the 
connected Green functions with 8 or more external legs are given by tree 
graphs made from $\Four$ and $\Ga^{(6)}$. 

If one wants to count loops instead of powers of the renormalized coupling function,
this system of equations is exact to two--loop order. However, we believe it is
more natural to think of an expansion in $\Four$ than a loop expansion. 

\begin{figure}
\begin{center}
\epsfig{file=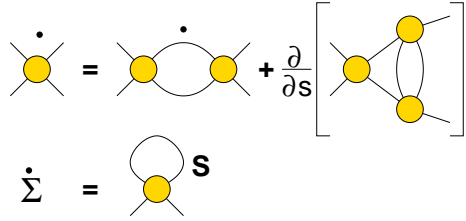,width=0.5\textwidth}
\end{center}
\caption{The graphical representation of the third order flow equations.
The dot denotes the derivative with respect to the scale parameter $s$.
All lines in the graphs carry full propagators $G$, except for the one
with a dot, which carries a $\dot G$, and the tadpole line in the equation 
for $\dot \Si$, which carries a single--scale propagator $S$.}
\label{fig0}
\end{figure}

The action of $\Lap_{S}$ can be described graphically as contracting 
any two of the external lines corresponding to a factor $\Psi_Y$ to an 
internal line, thus reducing the number of external legs to four. 
Terms where both derivatives of a Laplacian act on the same vertex
factor $\Four$ correspond to tadpole diagrams. If the derivatives
act on different vertex functions, we get a line of the graph connecting
these vertices. This gives the graphical representation of 
\Ref{eq:3Four} and \Ref{eq:3Si} depicted in Figure \ref{fig0}.

An important feature of \Ref{eq:3Four} is that the product of propagators
$G$ is differentiated. This will make it possible to integrate the flow equation
when only particular quadratic terms are kept on the right hand side. 
This corresponds to graphical resummations -- the RPA if only the particle--hole
RPA--type graph is kept, and the BCS ladder summation if the particle--particle
terms are kept. A similar structure appears in the Wick ordered form of the 
RGDE (see \cite{msbook}, Section 4.5.4 and 4.5.5). However, in 
\Ref{eq:3Four} 
the selfenergy is taken into account, which was not the case in the 
corresponding formulas of the Wick ordered flow. 

\subsection{Self--consistency}
In this section we illustrate how such a resummation works at the example
of the RPA plus Hartree resummation, which was originally given by 
Katanin \cite{Katanin}.

We drop the third order term from \Ref{eq:3Four}.
The remaining quadratic term in $\Four$ is identical to that in 
(52) of \cite{SalHon} except that the propagators $S$ and $G$ are replaced
by an $s$--derivative of two propagators $G$. Thus all considerations about
symmetries and the derivation of the equations for the coefficient functions
remain unchanged, and the equations (88)--(91) of \cite{SalHon},
as well as their graphical representation in Figures 4 and 5, 
carry over unchanged. The only replacement is
that instead of (92) we now have
\begin{equation}\label{eq:Lpm}
\cL_{\pm} (q,k) = \frac{\del}{\del s} \left( G(k) G(q\pm k) \right) .
\end{equation}
The RPA plus Hartree resummation corresponds to keeping only the 
RPA type particle--hole graph in Figure 4 (c)  and only the Hartree graph
in Figure 5 of \cite{SalHon}. This corresponds to the equations
\begin{eqnarray}
\dot V(p_1,p_2,p_3) 
&=&
\int \dd k \; \cL_{+} (p_2-p_3,k)
V(k,p_2,p_3) \;V(p_1,p_2-p_3+k,k) 
\label{eq:RPA}
\\
\dot \Si (p)
&=&
\int \dd k \;
V(p,k,k) S(k) 
\label{eq:Hartree}
\end{eqnarray}
The initial condition is $V(p_1,p_2,p_3) = V_0(p_1,p_2,p_3)$ at $s=0$.
Eq.\ \Ref{eq:RPA} has a unique solution, which satisfies  
\begin{eqnarray}\label{eq:RPArec}
V(p_1,p_2,p_3)
&=&
V_0(p_1,p_2,p_3)
\\
&+&
\int \dd k\;
V_0(p_1,k+ p_2-p_3,k)\; G(k) G(k+ p_2-p_3)\; V(k,p_2,p_3) ,
\nonumber
\end{eqnarray}
as can be verified by differentiating with respect to $s$. 
Iteration of \Ref{eq:RPArec} gives the RPA series for $V$. 
The selfenergy becomes 
\begin{eqnarray}
\dot \Si (p)
&=&
\int \dd k \;
V_0(p,k,k) S(k) 
\nonumber\\
&+&
\int \dd k\; V_0(p,k,k) G(k)^2 
\int \dd l \; V(k,l,l) S(l) .
\end{eqnarray}
The last integral over $l$ gives $\dot \Si(k)$. By \Ref{eq:dotG}, 
$G \dot \Si G = \dot G - S$, so 
\begin{equation}\label{eq:Hartreeproof}
\dot \Si (p)
=
\int \dd k \;
V_0(p,k,k) \dot G(k)  .
\end{equation}
With the initial condition $\Si =0$ for $s=0$ this integrates to 
\begin{equation}
\Si(p) = \int \dd k \;
V_0(p,k,k) G(k)  ,
\end{equation}
which is the Hartree equation for $\Si$ because $G= (Q-\Si)^{-1}$. 

\begin{figure}
\begin{center}
\epsfig{file=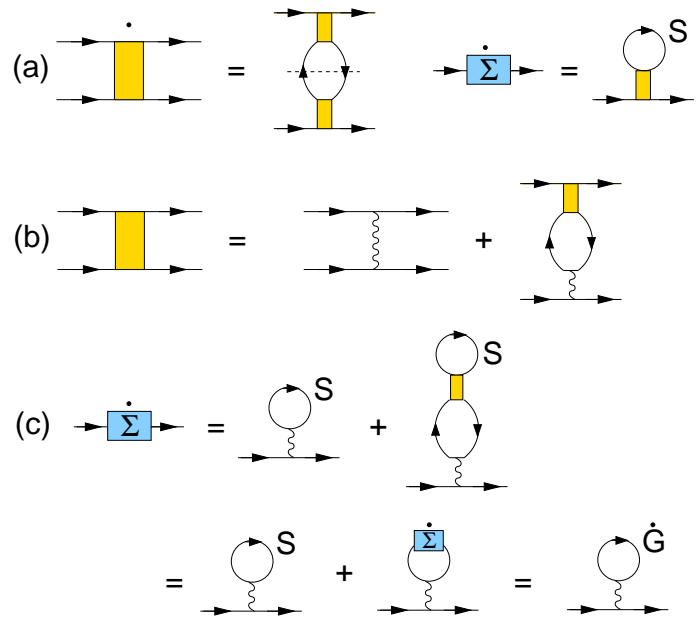,width=0.75\textwidth}
\end{center}
\caption{The RPA $+$ Hartee resummation. (a) The flow equations
\Ref{eq:RPA} and \Ref{eq:Hartree}. The dashed line indicates the 
$s$--derivative in \Ref{eq:Lpm}.
(b) The self--consistency equation for $V$ corresponding to 
Eq.\ \Ref{eq:RPArec}. 
(c) The graphical representation of the proof leading to Eq.\ \Ref{eq:Hartreeproof}: in the first line, (b) is inserted 
into the graph for the selfenergy from (a). In the second line,
the inner tadpole graph is replaced by the derivative of the selfenergy
by (a). Finally, the graphical relation corresponding to \Ref{eq:dotG} is used.}
\label{fig1}
\end{figure}

The diagrammatic form of this argument is shown in Figure \ref{fig1}. 

It is obvious that a similar argument works for the summation 
of the crossed particle--hole ladders and the Fock term for the 
selfenergy. If we resum particle--particle ladders, the Fock 
term vanishes unless we pose a small symmetry--breaking initial condition
for the selfenergy. In the next section we show that this reproduces
BCS theory. 

\section{Superconductivity} \label{sec:sc}

We now use the flow equations \Ref{eq:3Four} and \Ref{eq:3Si}
to derive an RG flow where an arbitrarily small symmetry--breaking field
is put in as an initial condition, and we show that this
produces a flow to a symmetry--broken state. 
In the standard Wilsonian picture for fixed points, 
a symmetry-breaking field moves the starting point  
in a relevant direction, and thus changes the fixed point 
at which the flow ends. 

\subsection{A small initial superconducting gap}
The role of the external field that breaks the symmetry 
in magnets, and which is sent to zero to study spontaneous 
symmetry breaking, is a small external field coupling to 
Cooper pair fields. In the fermionic action, it appears as 
a gap parameter $\Delta_0$. 

In our functional integral representation of the many--fermion model, 
the fermionic fields $\ps_\alpha (k) $ and $\psq_\alpha (k)$ 
depend on the momentum $k=(\omega,\vec{k})$,
where $\omega$ is the fermionic Matsubara
frequency and $\vec{k}$ the spatial part of the momentum,
and  the spin index $\alpha=\pm$. With the notation
$\int dk\; F(k) = T \sum_\omega L^{-d} \sum_{\vec{k}} F(\omega,\vec{k})$,
the quadratic part of the action is
\begin{eqnarray}\label{one}
\cA_0
&=&
\int \dd k\;
\big(
\psq(k) (\I \omega - e(\vec k)) \ps(k) 
\nonumber\\
&+&
\bar\Delta_0(\vec k) \psq(k) \frac{\veps}{2} \psq(-k)
-
\Delta_0(\vec k) \ps(k) \frac{\veps}{2} \ps(-k)
\big) .
\end{eqnarray}
Here 
$\psq(k)\ps(k) = \psq_+(k)\ps_+(k) +  \psq_-(k)\ps_-(k)$,
$e(\vec{k}) = \epsilon(\vec{k}) - \mu$ is the dispersion relation,
and 
$\psq(k) \veps \psq(-k) = \sum_{\alpha,\alpha'}
\psq_\alpha (k) \veps_{\alpha\alpha'} \psq_{\alpha'} (-k)$
with
$\veps_{+-}= - \veps_{-+} = 1$ and $\veps_{++} = \veps_{--} =0$.
The first term in $\cA_0$ is the usual Fermi gas kinetic term.
The dispersion relation $e$ includes the chemical potential $\mu$. 
The second contains a gap function $\Delta_0(\vec k)$ which acts as an 
external field coupling to the Cooper pairs. 
The Cooper pair is a singlet, so  $\Delta_0 (- \vec k) = \Delta_0(\vec k)$. 
We also assume that $\Delta_0$ depends only on $\vec{k}$.
A nonzero $\Delta_0$ breaks the charge symmetry. 
As $\Delta_0 \to 0$, the charge symmetry is restored in the action.
The question we take up here is what happens with expectation values
of a model where we add an attractive interaction of the fermions. 
Spontaneous symmetry breaking means that the limit of 
$\langle \psq\veps\psq\rangle $ is nonvanishing in the limit
$\Delta_0 \to 0$. 

To make contact with the formulas derived in the previous sections,
we introduce a Nambu--like field $\Psi$ as the column vector
\begin{equation}
\Psi(\omega,k) = (\psq_+(k), \psq_-(-k),\ps_+ (k), \ps_-(-k))^T
\end{equation}
(with the $T$ indicating the transpose) and 
rewrite the action as $\cA_0 = \frac12 (\Psi, \bigQ \Psi)$
where the brackets denote the bilinear form
$(f,g) = \sum_{j=1}^4 \int \dd k\; f_j (k) g_j(k)$,
with $j$ indexing the four components of $\Psi$, and the matrix
\begin{equation}
\bigQ (\omega, k) =
\left[
\begin{array}{cccc}
0 & \bar \Delta_0(\vec k) & \I \omega - e(\vec{k}) & 0 \\
-\bar \Delta_0(\vec k) & 0 & 0 & -\I \omega -e(-\vec{k}) \\
-\I \omega + e(\vec{k}) & 0 & 0 & - \Delta_0(\vec k) \\
0 & \I \omega + e( -\vec{k}) & \Delta_0(\vec k) & 0 
\end{array}
\right] .
\end{equation}
Note that our conventions differ from standard Nambu notation
in that the usual off--diagonal terms appear in the block--diagonal. 
This is because we want to have an antisymmetric matrix, corresponding
to the general formulas given in Section \ref{sec:1pi}.
We have also changed notation slightly, denoting by $\bigQ$ what was
called $Q$ in Section \ref{sec:1pi}. 

We are only interested in dispersion relations $e$ satisfying 
$e(-\vec{k}) = e(\vec{k})$, in which case
we can write $\bigQ$ in the block form 
\begin{equation}\label{eq:bQdef}
\bigQ (\omega, k) = 
\left[
\begin{array}{cc}
\veps \bar \Delta_0(\vec k) & \lQ(k) \\
-\lQ(k) & -\veps \Delta_0(\vec k)
\end{array}
\right]
\end{equation}
with 
\begin{equation}\label{Qdef}
\lQ(k) = \I \omega \Pauli_3 - e(\vec{k}) ,
\end{equation}
$\veps$ as above, and $\Pauli_3=$ diag $\{1,-1\}$. 

The fields in position space are defined as 
\begin{equation}
\check\Psi_j (x) = \int \dd k\; 
\E^{\I v_j k\cdot x} 
\Psi_j (k)
\end{equation}
with $x=(\tau,\vec{x})$ and $k\cdot x = - \omega \tau + \vec{k} \cdot \vec{x}$.
We take $(v_1,\ldots, v_4) = (-1,1,1,-1)$ to get the convention that $\psq $ 
and $\ps$ Fourier transform with opposite signs in the exponent.
This is necessary for the kinetic term to be of the form in \Ref{one}
because the bilinear form $(\cdot,\cdot)$ does not involve a complex conjugate.

\subsection{The RG flow for the reduced BCS model}
The RG scale $\Esc{s}= \Esc{0} \; \E^{-s}$ is used as an infrared cutoff.
That is, we replace $\lQ (k)$ in \Ref{eq:bQdef} by 
$\lQ (s,k) = \lQ(k) \chi(s, k)^{-1}$, where $\chi (s, k)$ is 
a smooth function that becomes very small when 
$| \lQ(k) | \le \Esc{s}$ and vanishes for $ \lQ(k)  =0$,
and $\chi (s,k) $ gets close to $1$ for $| \lQ(k) | \gg \Esc{s}$
(for the diagonal matrix $\lQ(k)$, $|\lQ(k)| = |\I \omega - e(\vec k)|$; 
in general one would take the smallest eigenvalue of $\lQ(k)$).
For the proof given in this Section 
that the gap flow leads to the exact solution of the BCS model
we do not need to assume a specific form of $\chi$, 
but only that $\chi (s,k) \to 1$ for $s \to \infty$. 
In particular, $\chi $ may also depend on the frequency $\omega$
(recall that $k=(\omega, \vec k)$). 
For the specific calculations in Section \ref{sec:gf}, 
we shall take a frequency-independent cutoff 
function and then take the limit of a sharp cutoff. 

In the RG flow, the gap function becomes a function $\Delta (s)$ of the 
scale parameter $s$.
If the interaction is attractive, $\Delta (s)$ increases with $s$, 
that is, as the energy scale $\Esc{s}$ is lowered.

We now derive the flow equations for the reduced BCS model
directly from the functional form of the equations 
\Ref{eq:3Four} and \Ref{eq:3Si}. We also show that the third order
term in \Ref{eq:3Four} is indeed irrelevant 
in this model in the thermodynamic limit. 
The essential reason for this
is that the special structure of the BCS interaction fixes loop variables
in this third order contribution. 
 
\subsection{The reduced BCS model}
The Hamiltonian of the usual reduced BCS model is of the form of 
a free fermion Hamiltonian $H_0$ plus a mean-field interaction
\begin{equation}\label{eq:redHam}
H=H_0 - g_0 L^{-d} C^\dagger C 
, \qquad\qquad 
C = L^{-d} \sum_{\vec k} f(\vec k) c_{\vec k,+}c_{-\vec k,-}
\end{equation}
The fermion operators $c_{\vec k,\al}$ have commutation relations
$\{c^{\phantom{+}}_{\vec k,\al}, c^+_{\vec k',\al'}\} 
= L^d \de_{\al,\al'} \de_{\vec k, \vec k'}$. 
The function $f(\vec k)$ is the gap symmetry function.
We have assumed singlet pairing, so $f(- \vec k) = f(\vec k)$. 
We assume that $g_0>0$ so that the interaction 
is attractive. 
Although the reduced interaction contains one momentum sum less 
than a more realistic short--range interaction, the inverse volume 
in \Ref{eq:redHam} is kept. 
This implies that in the corresponding functional integral, the 
frequency dependence of the interaction gets reduced to zero frequency pairs
in the thermodynamic limit \cite{Muehlschlegel}, so that the interaction 
effectively becomes 
\begin{equation}
S_{\rm rBCS} 
=
- \frac{g_0}{\Omega} \Xq \X
\end{equation}
with $\Omega = \beta L^d$ and 
\begin{eqnarray}\label{eq:rBCSaction}
\Xq &=&
-\frac{1}{\Omega}\sum_{\om,k}
\ovl{f} (\vec k) \,\psq(\om,\vec k)\,\frac{\veps}{2}\,
\psq (-\om,-\vec k) ,
\nonumber\\
\X
&=&
\frac{1}{\Omega}\sum_{\om,k}
f (\vec k) \,\ps(\om,\vec k)\,\frac{\veps}{2}\,
\ps (-\om,-\vec k) .
\end{eqnarray}
Again $\veps = \I \si_2$.
The functional integral with interaction \Ref{eq:rBCSaction} can be solved
explicitly in the sense that it can be reduced to an integral over a single 
complex variable. 
In the following, we calculate the RG flow for this model. 

The propagator in the flow equation is taken as $G=\bigQ^{-1}$ with $\bigQ $ 
given in \Ref{eq:bQdef}, but where $\De_0$ gets replaced by a scale dependent 
gap $\De(s)$. 
For the interaction, we make the ansatz
\begin{equation}\label{eq:Vansatz}
V(\Psi) 
=
-\frac{1}{\Omega}
\left(
v \Xq\X + \frac{w}{2} \X \X + \frac{\bar w}{2} \Xq \Xq
\right)
+ 
O \left( (\Omega)^{-2}\right)
\end{equation}
where $\Omega = \beta L^{d}$.
The $\X\X$ and $\Xq\Xq$ terms correspond to the non--charge--invariant parts of 
the vertex. Consequently, the initial condition for the vertices 
$v=v(s)$ and $w=w(s)$ is $v(0) =g_0$ and $w(0) = 0$. 
The coefficient of $\Xq\Xq$ is fixed to be the complex conjugate of $w$
to have $U(1)$ invariance of the initial interaction. 

The RG equations for $v$, $w$ and $\De(k)$ can now be straightforwardly
obtained from \Ref{eq:3Four} and \Ref{eq:3Si} by calculating the
action of the Laplacians $\Lap_G$. 
This is done in Appendix \ref{ap:Lapcalcs}. 
In that appendix, we also show that our ansatz, in particular 
\Ref{eq:Vansatz},  is complete for the 
reduced BCS model. That is, $\Omega$ plays a role analogous
to $N$ in a large--$N$ expansion, and the normal part of the 
selfenergy and any quartic term with a structure differing from 
that in \Ref{eq:Vansatz} gets an additional factor $1/\Omega$. 
In particular, the third order term in \Ref{eq:3Four} 
only produces contributions that are of order $\Omega ^{-1}$ 
compared to the terms in \Ref{eq:Vansatz}. 

It turns out (not surprisingly) that the $\vec k$--dependence of $\De$ 
must be $\De(\vec k) = f(\vec k) \de$ with a scale--dependent $\de=\de(s)$
and the function $f$ given in the reduced BCS interaction.
The initial condition for $\de $ is $\de(0)=\de_0$ with a nonzero $\de_0$.
We take $\de_0 >0$ to fix the phase of the gap. 
Then $w$ is real and the flow runs in the three--dimensional
space $(v,w,\de)$. The flow equations are
\begin{eqnarray}\label{eq:BCSflo1}
\dot v
&=&
\dot B (v^2 + w^2) + 2 \dot A vw
\\
\label{eq:BCSflo2a}
\dot w 
&=&
\dot A(v^2+w^2) + 2 \dot B vw
\\
\label{eq:BCSflo3a}
\dot \de 
&=&
C (v+w)
\end{eqnarray}
with 
\begin{eqnarray}
A
&=&
-
\frac{1}{\be}\sum_\om
\int \dd \vec k \; f(\vec k)^2 a_G(\om,\vec k)^2
\\
B
&=&
\frac{1}{\be}\sum_\om
\int \dd \vec k \; |f(\vec k)|^2 (\om^2+e(\vec k)^2) b_G(\om,\vec k)^2
\\
C
&=&
\frac{1}{\be}\sum_\om
\int \dd \vec k \; f(\vec k) a_S(\om,\vec k) 
\end{eqnarray}
and (denoting $|\sQ|^2 (\om,\vec k) = \om^2+e(\vec k)^2$)
\begin{eqnarray}
a_G
&=&
\frac{\De \chi^2}{|\sQ|^2 + (\De \chi)^2}
\\
b_G
&=&
\frac{\chi}{|\sQ|^2 + (\De \chi)^2}
\\
a_S
&=&
\De
\frac{2\chi\dot\chi |\sQ|^2}{[{|\sQ|^2 + (\De \chi)^2}]^2} .
\end{eqnarray}
The term $A$ is the particle--particle bubble with anomalous propagators,
$B$ is the particle--particle bubble with normal propagators, and 
$C$ is the loop integral corresponding to the Fock term
with an anomalous single--scale propagator. 

The equations for $g= v + w$ and $\th = v - w $ decouple:
\begin{eqnarray}
\dot g 
&=& 
(\dot A+\dot B) g^2 
\\
\dot \th 
&=&
(\dot B -\dot A) \th^2 
\end{eqnarray}
so their solution is
\begin{equation}\label{eq:gthflow}
g(s)=\frac{g(0)}{1-(A+B) g(0)} 
\qquad
\th(s)=\frac{\th(0)}{1-(B-A) \th(0)} .
\end{equation}
Because the initial interaction is charge invariant, 
$g(0)=\th(0)=g_0$.
Only the combination $g=v+w$ appears in the equation for the gap.
By \Ref{eq:gthflow}, the solutions for $v$ and $w$ obey the equations
\begin{equation}\label{eq:vwsols}
v 
=
g_0 + g_0 ( B v + A w) 
\qquad
w
=
g_0 (A v + B w).
\end{equation}
We can now show that the solution $\de$ of the flow for the gap
satisfies the BCS gap equation in the limit $s \to \infty$, where the 
RG scale $\Esc{0} \E^{-s}$ vanishes. Taking a derivative of $a_G$ with 
respect to the scale parameter $s$ we get the equation corresponding 
to  \Ref{eq:dotG},
\begin{equation}\label{eq:goodaS}
a_S  
=
a_{\dot G}  
-
\dot \De \; 
\left( |\sQ|^2 b_G^2 - a_G^2\right) .
\end{equation}
Thus
\begin{equation}
C
=
\frac{1}{\be}\sum_\om
\int \dd \vec k \; f(\vec k)^2 a_{\dot G} (k) \;
- \dot \De \;  (A+B) 
\end{equation}
and
\begin{equation}
\dot \De ( 1 + g (A+B))
=
g
\frac{1}{\be}\sum_\om
\int \dd \vec k \; f(\vec k)^2 a_{\dot G} (k)  .
\end{equation}
Since the solution for $g$ obeys
\begin{equation}
g_0 = \frac{g}{1+(A+B)g},
\end{equation}
we see that the equation for the gap becomes
\begin{equation}
\dot \de 
=
g_0
\frac{1}{\be}\sum_\om
\int \dd \vec k \; f(\vec k)^2 a_{\dot G} (k) 
\end{equation}
which integrates to 
\begin{equation}
\de(s) - \de_0 = 
g_0
\frac{1}{\be}\sum_\om
\int \dd \vec k \; f(\vec k)^2 
\frac{\de(s) \chi_s(\vec k)}%
{\om^2 + e(\vec k)^2 + (\de(s) f(\vec k) \chi_s(\vec k))^2}
\end{equation}
Here we have made explicit the $s$--dependence in the notation. 
In the limit $s \to \infty$, $\chi \to 1$ and we obtain 
\begin{equation}\label{eq:dezgap}
\de - \de_0 = 
g_0
\frac{1}{\be}\sum_\om
\int \dd \vec k \; f(\vec k)^2 
\frac{\de }{\om^2 + e(\vec k)^2 + (\de f(\vec k) )^2}
\end{equation}
For $\de_0 \to 0$ this becomes the BCS gap equation for $\de$:
\begin{equation}\label{eq:gapeq}
\de = 
g_0
\int \dd \vec k \; f(\vec k)^2 
\frac{\de }{2\sqrt{\de^2 f(\vec k)^2 + e(\vec k)^2}}
\tanh \left(
\frac{\beta}{2}
\sqrt{\de^2 f(\vec k)^2 + e(\vec k)^2}
\right)
\end{equation}
The limiting procedure $\de_0 \to 0$ picks the positive solution 
of this equation. The choice of cutoff function $\chi$ did not play
any role in this argument; this is an example for universality.

\subsection{The graphical representation of the argument}
\begin{figure}
\begin{center}
\epsfig{file=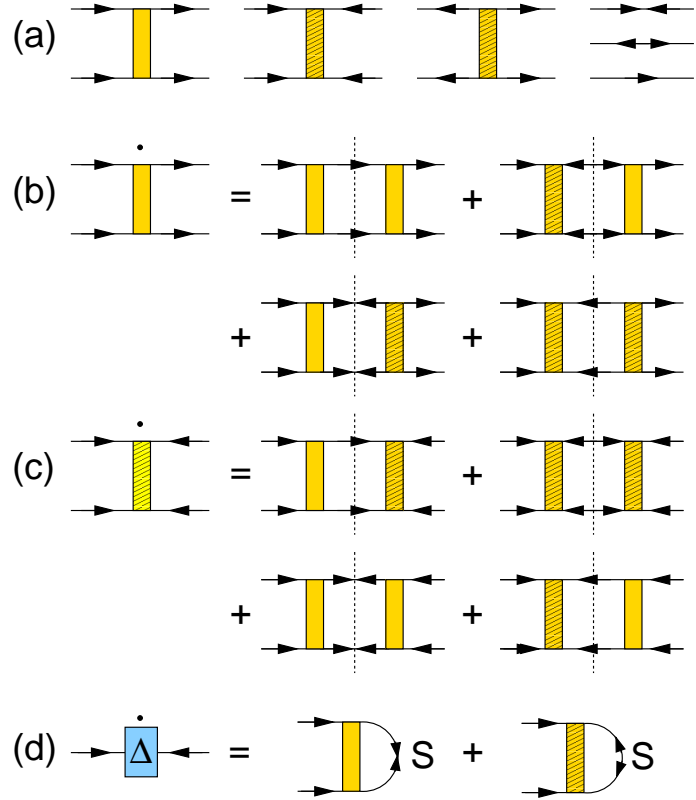, width=0.75\textwidth}
\end{center}
\caption{(a) The vertices and propagators in the superconducting gap flow.
(b),(c) RG flow equations for the normal vertex $v$ (Eq. \ref{eq:BCSflo1})
and anomalous vertex $w$ (Eq. \ref{eq:BCSflo2a}), respectively. 
The dashed line indicates that the derivatives of $A$ and $B$ with respect
to $s$ appear in the flow equations. 
(d) RG equation for the off-diagonal selfenergy (Eq. \ref{eq:BCSflo3a}).}
\label{fig2}
\end{figure}

\begin{figure}
\begin{center}
\epsfig{file=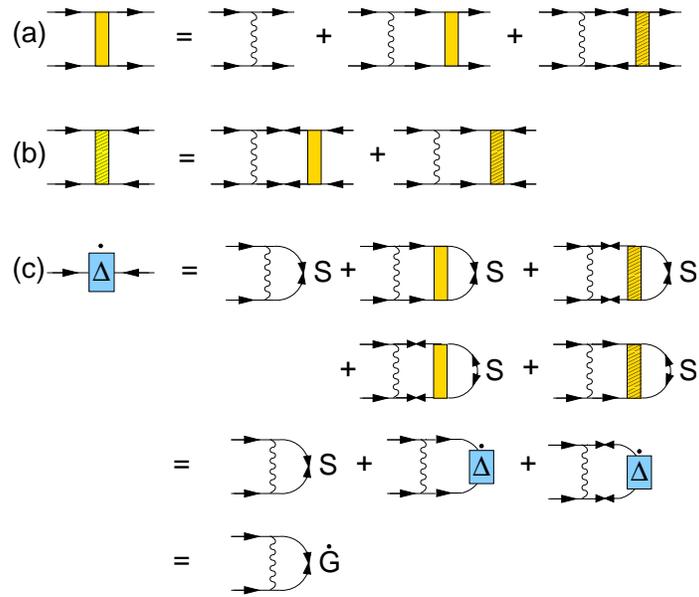, width=0.75\textwidth}
\end{center}
\caption{The graphical proof that the RG solution satisfies the BCS gap equation.
(a),(b) self-consistent equations for the normal vertex $v$ and
anomalous vertex $w$, respectively, corresponding to Eqns. \Ref{eq:vwsols}.
(c) RG equation \Ref{eq:BCSflo3a} for the off-diagonal selfenergy when (a) and
(b) are inserted as solutions for $v$ and $w$. This yields the
derivative of the usual BCS gap equation.}
\label{fig3}
\end{figure}

In Figures \ref{fig2} and \ref{fig3}, we show the vertices and flow equations
and exhibit the proof given in the previous subsection
that they resum ladders and anomalous Fock diagrams
leading to the BCS equation in graphical form. 
The graphical argument is closely similar to that depicted in Figure \ref{fig1}
except that there are now two types of interaction vertices,
and that the relation between $S$ and $\dot G$ involves a matrix, 
\begin{equation}
\dot G = S - G 
\left(
\begin{array}{rr}
\veps \dot{\bar \De} & 0 \\
0 & - \veps \dot \De
\end{array}
\right) G .
\end{equation}
Graphically, $\bar \Delta$ corresponds to a two--legged object 
with two outgoing arrows and $\De$ to one with two incoming arrows. 
For this reason, there are two $\dot \De$ terms in the 
next--to--last line of figure \ref{fig3}.
One can now write down the equations corresponding to these graphs
directly from the 1PI graph rules discussed
in \cite{SalHon}. 

\section{The gap flow} \label{sec:gf}
It is very instructive to look at the solution to the gap equation 
explicitly because one sees how the functions develop when 
$s$ varies. From this one can read off general qualitative properties
which we discuss below. 

\subsection{Specific choice of cutoffs and the resulting flow equations}
For the explicit representation of the resulting integrals 
it is convenient to take the cutoff function independent of $\omega$.
Then the Matsubara sums can be done explicitly. 
A possible choice is 
$
\chi(s,\vec k) = 
1 - 
{f_\ga (\sfrac{e(\vec k)^2}{\Esc{s}^2} -1)}{f_\ga (-1)}^{-1}
$
with some $\gamma \gg 1$ and $f_\ga (x) = (1+ \E^{\ga x})^{-1}$. 
We also take the limit of a sharp cutoff, $\ga \to \infty$, 
where $\chi$ becomes a step function (for a careful discussion
of how this limit is taken, see e.g.\ \cite{HonSalZ}). 
Then $\chi^2=\chi$, the term $\De \chi$ in the denominators in 
$A,B$ and $C$ simply becomes $\De$, and the 
single scale propagator becomes 
\begin{equation}
\sum_{\si=\pm} \de (\Esc{s}^{-1} e(\vec k) - \si )\;
\frac{\De}{\om^2 + e(\vec k)^2 + \De^2} .
\end{equation}
The flow equations then read 
\begin{eqnarray}\label{eq:BCSflo2}
\dot \de
&=&
g \; \de \;
\left\langle
f^2\;
\frac{\Esc{s} }{2\sqrt{\Esc{s}^2+\de^2f^2}}
\tanh \frac{\be}{2} \sqrt{\Esc{s}^2+\de^2f^2}
\right\rangle_{\Esc{s}}
\\
\dot g
&=&
g^2 
\;
\frac{\del}{\del s}
\left\langle
f^2\;
\left(
\frac{1}{2\;\cE }
\tanh \frac{\be}{2} \cE
+ 2 \de^2 f^2 F(\cE)
\right)
\right\rangle_{\ge \Esc{s}}
\\
\dot \th
&=&
\th^2 
\;
\frac{\del}{\del s}
\left\langle
f^2 \;
\frac{1}{2\;\cE }
\tanh \frac{\be}{2} \cE 
\right\rangle_{\ge \Esc{s}}
\end{eqnarray}
Here $\cE(\vec k) = \sqrt{e(\vec k)^2 + \de^2 f(\vec k)^2}$,
\begin{equation}
F(E)
=
\frac{1}{2 E }
\left[
\frac{\del}{\del E} 
\left(
\frac{1}{2E} \; \tanh \frac{\be}{2} E
\right)
\right] ,
\end{equation}
$\langle \;\cdot\; \rangle_{\Esc{s}} $ denotes the average 
on a single scale
\begin{equation}
\left\langle
A\right\rangle_{\Esc{s}}
=
\sum_{\si=\pm} 
\int \frac{\dd^d \vec k}{(2\pi)^d}
\;\de(e(\vec k) - \si \Esc{s})
\;
A(\vec k) ,
\end{equation}
and $\langle \cdot \rangle_{\ge \Esc{s}} $ denotes the average
above scale $\Esc{s}$, 
\begin{equation}
\left\langle
A\right\rangle_{\ge \Esc{s}}
=
\int\limits_{\vec k: |e(\vec k)| \ge \Esc{s}}
\frac{\dd^d \vec k}{(2\pi)^d}
\;
A(\vec k) .
\end{equation}
For an $s$--wave gap, $f=1$ and we can therefore rewrite 
the equations using the density of states $N(E)$. Let  
\begin{equation}
\tilde N(E) = N(E) + N(-E) 
\end{equation}
then
\begin{eqnarray}\label{eq:BCSflo3}
\dot \de
&=&
g \; \de \; \; \tilde N (\Esc{s})
\frac{\Esc{s}}{2\sqrt{\Esc{s}^2+\de^2}}
\tanh \frac{\be}{2} \sqrt{\Esc{s}^2+\de^2}
\\
\dot g
&=&
g^2 \;
\frac{\del}{\del s}
\int_{\Esc{s}}^\infty \dd E \; \tilde N(E)
\left(
\frac{1}{2\cE }
\tanh \frac{\be}{2} \cE 
+ 2 \de ^2 F(\cE )
\right)
\\
\dot \th
&=&
\th^2 \;
\frac{\del}{\del s}
\int_{\Esc{s}}^\infty \dd E \; \tilde N(E)
\frac{1}{2\;\cE }
\tanh \frac{\be}{2} \cE 
\label{eq:sharpthflow}
\end{eqnarray}
where now $\cE = \sqrt{E^2+\de^2}$. 

\subsection{Discussion of the solution}
With the sharp cutoff, \Ref{eq:gapeq} only gets a restriction 
$|E| \ge \Esc{s}$  in the integration for finite $s$, so that
\begin{equation}
\de(s) 
\left(
1-
g_0
\int_{\Esc{s}}^\infty \dd E \; \tilde N(E)
\frac{1}{2\;\cE }
\tanh \frac{\be}{2} \cE 
\right)
=
\de_0
\end{equation}
\null From this and \Ref{eq:sharpthflow}, we obtain the solution
\begin{equation}\label{eq:tangex}
\th(s) 
=
\frac{g_0}{1- g_0
\int_{\Esc{s}}^\infty \dd E \; \tilde N(E)
\frac{1}{2\;\cE }
\tanh \frac{\be}{2} \cE  }
=
g_0 \frac{\de(s)}{\de_0} 
\gtoas{s \to \infty}
g_0 \frac{\de}{\de_0}
\end{equation}
which is the exact solution for the tangential vertex. 
It diverges in the limit $\de_0 \to 0$; this is the remnant of
the $1/q^2$ singularity of the Goldstone boson in this reduced model. 
The solution for $g$ is
\begin{equation}
g(s)
=
\frac{1}{\th(s)^{-1} - 2 
\int_{\Esc{s}}^\infty \dd E \; \tilde N(E)\;
\de^2 F(\cE) }
\end{equation}
which for $s \to \infty$ converges to the exact solution for
the radial vertex. It stays finite in the limits $s\to \infty$ 
and $\De_0 \to 0$. This reflects the fact that amplitude fluctuations 
of the order parameter are gapped. 

\begin{figure}
\begin{center}
\includegraphics[width=.7\textwidth]{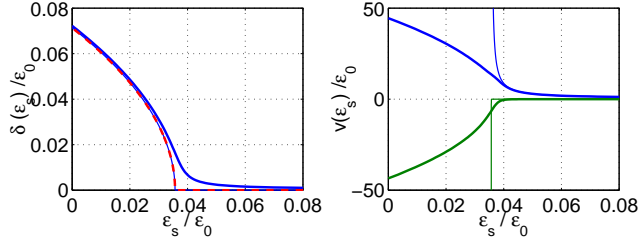}
\end{center}

\caption{Left plot: Gap amplitude $\delta (s) $ versus cutoff 
$\Esc{s} $ from the RG at zero temperature, density of states $N_0=1$,
and initial interaction $v_0/\Esc{0}=0.3$ in units of the bandwidth $\Esc{0}$. 
For $\delta_0=0$, the Cooper instability is at $\Esc{s}^c/\Esc{0}= 0.0357$.
The heavy line is for initial gap $\delta_0/\Esc{0}=2.4 \times 10^{-4}$ 
while the thin solid line is for $\delta_0/\Esc{0}= 6 \times 10^{-8}$. 
The dashed line (right on top of the thin solid line) is 
the asymptotic cutoff-dependence 
$\delta (\Esc{s}) = 2\sqrt{\Esc{s}^c (\Esc{s}^c - \Esc{s})}$ 
for $\Esc{s} < \Esc{s}^c$ and $\delta_0 \to 0$.  
Right plot: Flow of the normal vertex $v_s$ (upper curves) and anomalous vertex $w_{s}$ (lower curves) for the two different initial gaps. For $\Esc{s} \to 0$, $w_s\to -v_s$.
 }
\label{GFplot}
\end{figure}

\begin{figure}
\begin{center}
\includegraphics[width=.7\textwidth]{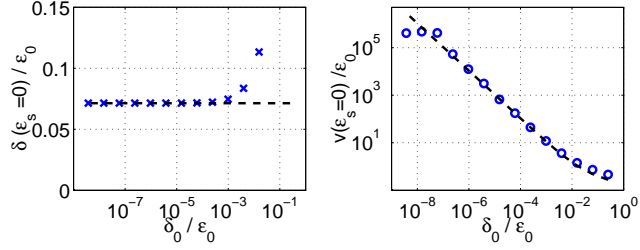}
\end{center}
\caption{Left plot: Final gap $\delta_{\rm final}$ ($s \to \infty$) versus initial gap $\delta_0$. the dashed line is the BCS value $\delta= 2\Esc{0}\exp (-1/N_0v_0)$ for $v_0=0.3$ and $N_0=1$. 
Right plot: Log-Log-plot of the final coupling strength  $v_{\rm final}$ versus initial gap $\delta_0$. The dashed line is the BCS result $v= v_0 \delta/(2\delta_0)$ for the vertex in presence of a symmetry-breaking external field $\delta_0$. The deviations at very small $\delta_0$ are numerical errors when the flow of the vertices becomes very steep.
}
\label{initialgaps}
\end{figure}

The solution of the equations is shown in Figures
\ref{GFplot} and \ref{initialgaps}.
Characteristically, the gap remains small of order $\De_0$ 
in the flow down to a scale 
$\Esc{s}^*$. Then it rapidly starts to grow
and it finally saturates at a value that is the closer to the 
BCS solution the smaller $\De_0$ was. 
For small $\De_0$, the scale $\Esc{s}^*$ is only a little bit larger
than the scale $\Esc{s}^c$ where the flow with $\De_0 = 0$ diverges.
For $\De_0=0$, the flow of the gap is identical to that of the 
superconducting susceptibility studied 
in \cite{ZS,HM,HSFR,SalHon}. 
For $\Delta_0 \ne 0$, the coupling constants become large, but they do not
diverge in the flow. 

The qualitative features of the flow for the gap are easily 
understood from the equation \Ref{eq:BCSflo3} itself. 
The right hand side is positive, so the gap grows with $s$. 
On the other hand, $\Esc{s}$ decreases with $s$, so there is 
a crossing scale where $\De(s) = \Esc{s}$. When $\De_{s}$
gets much larger than $\Esc{s}$, the ratio 
$\sfrac{\Esc{s}}{\sqrt{\Esc{s}^2+\de(s)^2}}$ 
in \Ref{eq:BCSflo3} becomes proportional to $\Esc{s}$ 
so the scale derivative of the gap vanishes and the 
gap eventually saturates. 
This is also true for gap functions $f$ that have nodes:
in this case, there is always a region of momentum space
where $\de(s) \le \Esc{s}$ but this region shrinks to 
the nodal points for $s\to \infty$ and its contribution 
to the right hand side of the flow equation vanishes 
for $s \to \infty$.  

The gap flow thus exhibits three regions of clearly 
distinct behavior, which also exist in non--reduced models.
Initially, the gap is much smaller than the energy scale,
so it affects the flow very little and the flow stays close
to that of the susceptibility. Indeed, as long as the gap
is small compared to the energy scale, an expansion of the 
propagators in the gap parameter is convergent because the
energy scale $\Esc{s}$ is an infrared cutoff on the propagators. 

In the final stages of the flow, the absolute value of the
gap is much larger than the energy scale. 
In a non--mean field model, the gap may still have 
phase fluctuations. 
If no vortices are present, a gauge transformation can 
be used to move the phase dependence from the gap to
the hopping term of the effective action, 
which is much smaller than the absolute 
value of the gap and can therefore be expanded in a convergent series. 
Thus now the gap term is the dominant term in the denominator
and it determines the long--distance behavior of the fermion 
propagator.
 
The intermediate regime is that where $\De(s)$ and $\Esc{s}$ cross,
i.e.\ the gap is comparable to the RG energy scale. 
In this regime one expects phase fluctuations to be
of great importance for the detailed behavior of the 
fermion propagator.


\subsection{Caveats}
\begin{figure}
\begin{center}
\includegraphics[width=.7\textwidth]{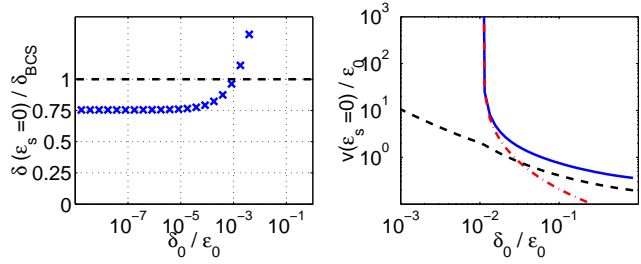}
\end{center}
\caption{Results in  the unmodified 1PI scheme. Left plot: Final gap
$\delta_{\rm final}$, normalized w.r.t. the BCS value, versus initial
gap $\delta_0$. The dashed line is the BCS value $\delta= 2\Esc{0}\exp
(-1/N_0 v_0)$ for $v_0=0.3$ and $N_0=1$.
Right plot: Log-Log-plot of the final coupling strength  $v_{\rm final}$
(solid line) and the anomalous vertex $ w_{\rm final}\times -1$
(dashed-dotted line)  versus initial gap $\delta_0$. The dashed line is
the BCS result $v= v_0 \delta/(2\delta_0)$ for the vertex in presence of
a symmetry-breaking external field $\delta_0$. For values 
of $\delta_0/\Esc{0}$ smaller than about $10^{-2}$, 
the integration diverges at a nonzero scale $\Esc{s}$.
}
\label{inigapwokat}
\end{figure}

As discussed, the general properties of the gap equation already 
imply that the gap saturates at low scales. 
However, they do not imply that the gap saturates {\em at the correct value}. 
This happens only because $\dot G$ was used instead of $S$ 
(the effect of that replacement on the vertex is even more drastic;
see below). Here we show how things go wrong when this
replacement is not made, i.e.\ we now consider the truncation 
to the second order flow that is obtained by discarding the 
1PI six--point function from the flow equations. 
The only change in the flow equation is that the $\dot G$
gets replaced by $S$. The rest of the structure remains
identical. 
A sample result of such a flow is shown in Figure \ref{inigapwokat}.
Although the gap grows qualitatively correctly in the flow 
and saturates at a value 
near the gap given by the BCS equation, it  is considerably smaller
than the correct solution to the gap equation.
Moreover, the two--particle vertex diverges at a nonzero scale
even at positive $\De_0$, in contradiction to the exact
solution \Ref{eq:tangex}.
The reason for this behavior is that putting $S$ on the propagator
lines instead of $\dot G$ neglects contributions to the gap
from higher scales. Because the gap term regularizes the propagator, 
the speed of the flow of the four--point vertex depends on the gap. 
Because the gap is too small, the vertex grows too fast and 
has an artificial divergence at a nonzero scale. This behavior 
can be understood in that the Ward identity from global $U(1)$ charge symmetry
is violated in this truncation. As mentioned, the replacement of $S$
by $\dot G$ leads to a rearrangement of the hierarchy that involves 
all orders; the correct BCS flow can be recovered only in this way. 

Finally let us mention that if only the $v$--vertex is kept, 
the final gap does saturate at small scales, but at a value 
that diverges in the limit $\De_0 \to 0$, so that this approximation only 
reproduces the flow  of the superconducting susceptibility 
in the limit $\De_0 \to 0$. 

\subsection{The Ward identity}
\label{se:WI}
The Ward identities corresponding to $U(1)$ gauge transformations 
of the many--fermion system are of central importance for the system,
especially in its transition to the symmetry-broken phase. 
The existence and essential properties of the Goldstone bosons 
arising from the spontaneous breaking of the symmetry are directly 
connected to the Ward identities. 

Satisfying the Ward identities is difficult in any approximate scheme. 
In the flow equation approach, finding truncations that lead to the correct
Ward identities at the end of the flow is a nontrivial problem, 
which was addressed in \cite{Katanin}. A general investigation of the 
role of Ward identities in functional RG flows is done in \cite{EnssDiss}.

In the reduced BCS model, the Ward identity is reduced to that corresponding 
to global gauge transformations. Although this is only a one--parameter 
subgroup, its preservation
is nontrivial in approximate schemes: the symmetry breaking relates the value of
the order parameter with a divergence of the two--particle vertex at scale zero.
We show explicitly below that this Ward identity is satisfied in the full propagator
scheme. It is violated in other truncations, in particular in the standard formulation 
of the 1PI flow, where $S$ is not replaced by $\dot G$. 
This is the deeper reason for the above--discussed
failure of the unmodified 1PI scheme 
in the symmetry--broken phase. 

\subsubsection{General structure of Ward identities}
Ward identities can be derived under very general conditions and for general symmetries. 
We assume that we have a linear invertible transformation $\SY$ of the fields, $\Psi \to \Psi' = \SY \Psi$
which leaves the initial interaction potential $V_0$ and the integration measure invariant, 
\begin{equation}
V_0(\SY \Psi ) = V_0(\Psi) \mbox{ and } \cD \left(\SY\Psi \right) = \cD \Psi .
\end{equation}
In general, such a symmetry transformation can mix the fields 
at different space--time points and act nontrivially on all internal 
indices, i.e.\ written in detail, it reads 
\begin{equation}
\Psi'  (X') = \int \dd X \; 
\SY (X',X) \, \Psi (X)
\end{equation}
A change of integration variable in \Ref{eq:Wdef} from $\Psi$
to $\Psi'$ implies
\begin{equation}\label{eq:WWI}
W(\SY^T Q \SY, \SY^T H) = W (Q,H) .
\end{equation}
Here we have again made the dependence of $W$ on the quadratic part $Q$ 
of the action in \Ref{eq:Wdef} explicit.
The Ward identity for $W$ can be generated directly from this equation.
It is already clear from this equation that the Ward identity has a form very similar
to that of the RG flow equation: the symmetry transformation also changes 
only the 
quadratic part $Q$ of the action. An infinitesimal change can be rewritten in terms of 
a Laplacian with respect to the source terms in the same way as was used in 
\cite{SalHon} to derive the flow equation. 
However, now the scale derivative of $Q$ is replaced by a commutator of $Q$ with
the symmetry transformation and an additional dilatation operator appears,
as will be explained now in more detail.  

So let us 
assume further that the group of symmetries $\SY$ is a Lie group and 
consider an infinitesimal transformation $\SY = 1 + \SG + O(\SG^2)$.
To linear order in $\SG $ we have 
\begin{equation}
- \left(
\SG  H , \sfrac{\de}{\de H} 
\right)\; W
+
\E^{W} \; \frac12\Lap_{\SG ^T Q + Q \SG } \; \E^{-W} 
= 0.
\end{equation}
For $\SG ^T=- \SG $, the operator in the Laplacian is a commutator. 
In general (and in our specific setting) one has commutators and anticommutators; see below.
Obviously, the form of this equation is similar to that of 
the flow equation \Ref{eq:WPolch} for $W$.
The reason for this similarity is that both the scale change and 
the symmetry transformation change only the quadratic term
of the action, not the interaction potential $V_0$. 
By the standard relations between $W$ and its Legendre transform $\Ga$,
we obtain the Ward identity for $\Ga$
\begin{equation}
- \left(
\SG \frac{\de \Ga}{\de \Psi}\, , \Psi
\right)
-
\frac12 \Tr 
\left( 
[Q,\SG ] 
\left(
\frac{\de^2 \Ga}{\de \Psi^2}
\right)^{-1}
\right)
+
\left(
\Psi\, ,
[Q,\SG ] \, \Psi 
\right)
=
0 .
\end{equation}
In the next subsection, we specialize to the identities we need for
our specific case of a spontaneously broken $U(1)$ symmetry.

\subsubsection{The $U(1)$ Ward identities for the superconducting state}
To make the effect of the symmetry--breaking terms explicit 
in the subsequent discussion,
we switch back from the Nambu--type field $\Psi$ 
to the fields $\ps$ and $\psq$. Thus 
\begin{equation}
Z(\etq,\et,Q,\bDe)
=
\int \cD \psq \cD\ps \; 
\E^{- (\psq,Q\ps) - \frac12 (\psq, \ovl{\bDe} \psq) + \frac12 (\ps, \bDe\ps)}
\E^{- V_0(\psq,\ps) + (\etq,\ps) + (\psq,\et)}
\end{equation}
A general gap term is given by 
\begin{equation}
(\ps, \bDe\ps)
=
\int \dd \xi \int \dd \xi'\;
\ps(\xi) \bDe (\xi,\xi') \ps(\xi')
\end{equation}
with an antisymmetric function $\bDe$ of $\xi = (x,\si)$ and $\xi'$. 
The gauge transformation is 
\begin{equation}
\ps_\si (x)  = \E^{\I \al(x)}\; \ps'_\si (x), 
\quad
\psq_\si (x)  = \E^{- \I \al(x)}\; \psq'_\si (x) .
\end{equation}
As before we get
\begin{equation}
Z(\etq,\et,Q,\bDe_0)
=
Z(\E^{\I \al} \etq, \E^{- \I \al} \et, \E^{- \I \al} Q \E^{\I \al},
\E^{\I \al} \bDe_0 \E^{\I \al} ).
\end{equation}
Taking an infinitesimal transformation, we get for $W = - \ln Z$
\begin{eqnarray}
0 
&=&
\left[
(\I \al \etq, \sfrac{\de}{\de \etq})
-
(\I \al \et, \sfrac{\de}{\de \et})
\right]
W
+ \E^W 
\left(
\sfrac{\de}{\de \et} , \I [\al,Q] \sfrac{\de}{\de \etq}
\right)
\E^{-W} 
\nonumber\\
&-&
\E^W 
\frac12
\left[
\left(
\sfrac{\de}{\de \etq}, 
\I \{ \al, \bDe_0 \}
\sfrac{\de}{\de \etq}
\right)
+
\left(
\sfrac{\de}{\de \et}, 
\I \{ \al, \ovl{\bDe_0} \}
\sfrac{\de}{\de \et}
\right)
\right]
\E^{-W} 
\end{eqnarray}
By Legendre transformation, 
this implies the following Ward identity for the 1PI generating function 
$\Ga(\phq,\ph)$
\begin{eqnarray}
0 
&=&
(\I \al \phq, \sfrac{\de \Ga}{\de \phq})
-
(\I \al \ph, \sfrac{\de \Ga}{\de \ph})
\nonumber\\
&-&
(\phq, \I [\al,Q] \ph) 
- \frac12
\left[
(\ph, \I \{ \al, \bDe_0 \} \ph) 
+
(\phq, \I \{ \al, \ovl{\bDe_0} \} \phq )
\right]
\\
&+&
\frac12
\mbox{ Tr }
\left[
\left(
\begin{array}{rr}
\I \{ \al, \ovl{\bDe_0} \} & \I [\al,Q] \\
- \I [\al,Q]^T & \I \{ \al, \bDe_0 \}
\end{array}
\right)
\left(
1+ G \tilde \Ga^{(\ge 4)}
\right)^{-1}
G
\right] .
\nonumber
\end{eqnarray}
Again there is some similarity with the flow equation, 
but now also off--diagonal terms appear, because the 
gauge transformation affects all quadratic terms in the action. 
Here 
\begin{equation}
G^{-1} =
\left(
\begin{array}{cc}
\bar \bDe  &  (Q-\Si) \\
- (Q-\Si)^T &  - {\bDe}
\end{array}
\right) .
\end{equation}
The usual Ward identities are now generated by expanding in the fields.
In second order in $\phi$ and $\phq$, 
only the four--point vertex $V=\Ga^{(4)}$ contributes and we obtain 
a relation between the normal and anomalous selfenergy and the 
corresponding parts of the vertex. We shall need here only the 
relation for the reduced BCS model and for a gauge transformation 
with constant $\al$. In that case $[\al,Q]=0$ and 
$\{ \al, \bDe_0 \} = 2 \al \bDe_0$. Moreover, 
for the singlet pairing considered here, 
$\bDe((x,\si),(x',\si')) = \veps_{\si,\si'} \check \De (x-x')$
where $\check\De$ has Fourier transform $\De (\vec k)= f(\vec k) \de $,
and in the thermodynamic limit
$\Si =0$.  Comparing the $\phi \phi$ term gives the relation
\begin{equation}
\de - \de_0 
=
\de_0 (v-w) \frac{1}{\be}\sum_\om
\int \dd \vec k \; f(\vec k)^2 
\frac{1}{\om^2 + e(\vec k)^2 + (\de f(\vec k) )^2}
\end{equation}
Together with the gap equation \Ref{eq:dezgap}, this implies
\begin{equation}
\frac{\de}{\de_0} = \frac{v-w}{g_0} 
\end{equation}
which is the exact solution for the tangential vertex $v-w$.
Thus, given $\De$, the vertex is 
determined by the Ward identity, 
and the exact result \Ref{eq:tangex}
for the vertex is recovered only if the
gap satisfies the correct gap equation.

\section{The gap flow for general models} \label{sec:genmod}
The reduced BCS model is  a mean--field model for which the 
resummation of the particle--particle ladders and the 
Fock equation for the anomalous selfenergy are exact. 
We have seen that these features also come out naturally
in our gap flow. If we had not introduced the initial gap, 
we would have found a flow to strong coupling with a divergence
of the coupling function at a scale just below the one where
the gap starts to grow in the flow. In this way our calculation
also provides an example how a fermionic flow to strong coupling
is converted to a regular flow to a symmetry--broken state. 
In our example, this state corresponds to the superconducting
fixed point of the RG. 

Our method also applies to more realistic short--range models.
In these cases, the interaction vertex includes more terms 
than in the reduced BCS model. 
Then the particle-hole diagrams will give nonzero contributions in the flow.
Moreover, there are anomalous
``3:1'' vertices with three particles going in and one coming out,
as well as others that vanish in the thermodynamic limit of 
the reduced BCS model. 
With these straightforward generalizations, the method can
be applied to any model that has superconducting correlations.

Because the gap flow is a smooth modification of the susceptibility 
flow it automatically generates the correct gap 
symmetry in the course of the flow -- wrong gap symmetries
remain suppressed in the gap flow just as they are in the 
susceptibility flow. 
One can also study flows where the shape function $f$ of the 
gap is not fixed but itself determined during the flow.  
Let us give a typical example. In the repulsive Hubbard model on the two-dimensional square lattice, the initial interaction does not contain any attractive pairing channels. However in the course of the RG flow,  particle-hole corrections generate an attractive $d_{x^2-y^2}$-wave component. Stopping the RG flow a certain scale  
and then resorting to a mean-field solution of the theory at lower scales 
can lead to ambiguities regarding the correct scale for joining these two treatments, and all interaction terms involving $\psi_k \psi_{k+q}$ with 
small $q \ne 0$ have to be discarded. 
The gap flow finds the order parameter without these difficulties. 

It is also possible to include more than one order parameter in the initial conditions, 
hence to study the competition of different ordering tendencies, 
such as antiferromagnetism and superconductivity.  
In summary, we believe that the method has the potential to become 
a useful and convenient calculational tool for systems with competing
interactions.

An alternative way of viewing the initial scale $\De_0$ of the 
gap flow is as follows. The initial interaction may contain
many other terms 
in addition to the interaction that, either directly or 
by higher order effects, generates superconductivity. 
Since the interaction in a real system 
is determined only up to some small error terms, one can 
imagine that some very small interactions are present that have
a nonzero component of a pairing interaction of the form
$g_1 \bar X X$, where $X$ and $\bar X$ are as in the reduced 
BCS model, with some gap symmetry function $f$. If we perform a 
Hubbard--Stratonovich transformation on these small terms, 
the partition function becomes 
\begin{equation}
\int \frac{\dd \bar h \wedge \dd h}{2\pi\I g_1}
\E^{-|h|^2/g_1}
\int \cD \Psi
\E^{\bar h X + h \bar X}
\E^{-S(\Psi)}
\end{equation}
Now the HS field $h$ plays the role of the initial gap $\De_0$,
and the fermionic integral at fixed $h$ can be done by  the RG flow.
The only difference is that 
instead of taking the limit $h \to 0$, we now have to integrate 
over $h$. The smaller $g_1$, the more concentrated the 
integral will be around $h=0$. The point $h=0$ itself leads to 
a divergent flow, but in the integral over $h$, this is inessential
because at any other value of $|h|$, the flow converges to a 
finite result, which gets closer to the BCS gap if $|h|$ is made
smaller (and if the gap symmetry is the correct one). 
In other words, the exceptional point $h=0$ is really
irrelevant by a measure zero argument. 

It is important to note that in this approach, 
there is still the overall integration 
over the phase $\vphi$ of $h$ which is part of the $h$ integral. 
It leads to a vanishing expectation value of $X$ even in the symmetry 
broken phase. This is easily understood: picking a particular $\vphi$
corresponds to selecting a particular pure phase. 
Taking the average over $\vphi$ produces a mixed phase
in which noninvariant expectation values vanish 
but off--diagonal long range order persists.

\begin{appendix}

\section{Derivation of the flow equations}\label{ap:Lapcalcs}
The full propagator is 
\begin{equation}
G = \bigQ^{-1}
=
\frac{\chi}{\om^2+e(\vec k)^2 + (\chi \De)^2}
\left(
\begin{array}{rr}
-\veps \De \chi & - \bar \lQ \\ \bar \lQ & \veps \bar \De \chi
\end{array} 
\right)
\end{equation}
The single-scale propagator is 
\begin{equation}
S
=
\frac{\dot\chi}{\om^2+e(\vec k)^2 + (\chi \De)^2}
\left(
\begin{array}{rr}
-2 \veps \De \chi \lQ \bar \lQ & B \\ -B  & 2 \bar \lQ \lQ \veps \bar \De \chi
\end{array} 
\right)
\end{equation}
where $B = \bar \lQ (|\De^2| \chi^2 - \lQ \bar \lQ) = B^T$. 

\subsection{The main terms}
Consider the term quadratic in $V$ on the RHS of \Ref{eq:3Four}. 
The square of the Laplacian removes four fields from the degree 8 polynomial
in the fields. We can group the terms in two categories, which we now discuss
at the example of one term in the product $V^{(1)}V^{(2)}$, namely 
\begin{equation}
T
=
\BEqual{
\Lap^2
\frac{w\bar w}{4} X^{(1)}X^{(1)} \bar X^{(2)} \bar X^{(2)}
}
\end{equation}
Here we abbreviated $\Lap=\Lap_G^{(1,2)}$ and $X(\Psi^{(i)}) = X^{(i)}$. 
The first type of terms is the one where two of the four factors remain
untouched by derivatives. There are four such terms. Calling the sum of all other terms
$\cR$, we have 
\begin{equation}
T=
w \bar w\;
\bar X \, X\; 
\BEqual{
\Lap^2 \left(\bar X^{(1)} X^{(2)}\right)
}
+ \cR
\end{equation}
$\cR $ gets contributions from the terms of the second type where derivatives
act on at least three factors. 
All terms contributing to $\cR$ vanish in the limit $L \to \infty$ because 
a loop integration gets fixed by an external momentum (this is specific 
to the reduced BCS interaction and tracing this effect gives a diagrammatic
argument that the reduced BCS model is exactly solvable). 
We illustrate it at one
example in the next subsection. Dropping these remainder terms $\cR$ 
in all contributions, calling 
\begin{eqnarray}
B &=&
\frac{1}{\Omega}
2 \Equal{
\Lap_G^2 (X^{(1)}\bar X^{(2)})
}
=
\frac{1}{\Omega}
2 \Equal{
\Lap_G^2 (\bar X^{(1)} X^{(2)})
}
\\
A
&=&
\frac{1}{\Omega}
2 \Equal{
\Lap_G^2 (X^{(1)} X^{(2)})
} ,
\end{eqnarray}
and using that 
\begin{equation}
\frac{1}{\Omega}
2
\Equal{
\Lap_G^2 (\bar X^{(1)} \bar X^{(2)})
}
=\bar A
\end{equation}
we get the equations
\begin{eqnarray}
\dot v 
&=&
\dot B (v^2 + w \bar w) + (\dot A v w + \dot{\bar A} v \bar w)
\\
\dot w
&=&
\dot{\bar A} v^2 + \dot A w^2 + 2 B vw
\\
\dot {\bar w}
&=&
\dot A v^2 + \dot{\bar A} {\bar w}^2 + 2 B v \bar w
\end{eqnarray}
If we assume $\De_0 >0$, $w=\bar w$ is real and these equations simplify to
\begin{eqnarray}
\dot v 
&=&
\dot B (v^2 + w^2) + 2 \dot A vw
\\
\dot w
&=&
\dot A(v^2+w^2) + 2 \dot B v w .
\end{eqnarray}
The equation for the scale--dependent gap is 
\begin{equation}
\dot \de
=
v \Lap_S \bar X + w \Lap_S X 
\end{equation}
(as in the coupling flow, the normal part of the 
selfenergy vanishes for $L \to \infty$ because a loop integration
gets fixed due to the special structure of the reduced BCS interaction,
and there remains an overall factor $L^{-d}$).
For $\De_0 > 0$ this again simplifies to 
\begin{equation}
\dot \De
=
g \Lap_S X
\end{equation}
The graphical interpretation of the coefficients $A,B$ and $C$ is straightforward:
$A$ and $B$ denote the particle--particle bubbles of the theory;
$A$ is the one with anomalous
terms and $B$ the one with normal terms. $C$ is the value of the Fock diagram
with the single--scale propagator $S$ on the line. 

\subsection{The remainder terms}
In this section, we first show two prototypical terms which 
turn out to be one power of $\Omega$ smaller than the main terms,
and hence vanishing in the limit $\Omega \to \infty$, relative to the 
main terms. After that, we give the general argument for the reduced 
BCS model. This will also make clear how the terms that combine with 
$S$ to form $\dot G$ are singled out in a natural way in this model, 
hence we give a specific and complete justification for 
the replacement of $S$ by $\dot G$ in the 1PI equations for this example. 

Recall that the four--point function is of order $1/\Omega$
in the reduced BCS model right from the start, so subleading corrections
have to be of order $1/\Omega^2$. The two--point function and the selfenergy
are of order $1$. That is, in the limit $\Omega \to \infty$, the model describes
generalized free fields. 

We start with a few preparations. In terms of the field components $\psq$ and $\ps$, 
the Laplacian reads
\begin{equation}
\Lap_G
=
- \left( \frac{\de}{\de \psq}, \; \bar \lQ b_G \frac{\de}{\de \ps}\right)
-\left( \frac{\de}{\de \psq}, \;  \veps a_G \frac{\de}{\de \psq}\right)
+\left( \frac{\de}{\de \ps}, \;  \veps a_G \frac{\de}{\de \ps}\right)
\end{equation}
and similarly for $S$. Here we used 
\begin{equation}
G 
=
\left(
\begin{array}{rr}
-\veps a_G & -\bar \lQ b_G \\
\bar \lQ b_G & \veps a_G \\
\end{array}
\right) .
\end{equation}
The functional derivatives obey
\begin{equation}
\frac{\de}{\de \ps_\al(k)} \ps_{\al'} (k') 
=
\de_{\al,\al'}
\;
\de_{\Om} (k,k')
\end{equation}
where 
$\de_{\Om} (k,k') = \be \de_{\om,\om'} \; L^d \de_{\vec k,\vec k'}$.
With this convention, one can take the limit $L \to \infty$ 
that we are interested in, as well as the limit $T \to 0$,
without any further rescaling of the fields. 

\begin{figure}
\begin{center}
\epsfig{file=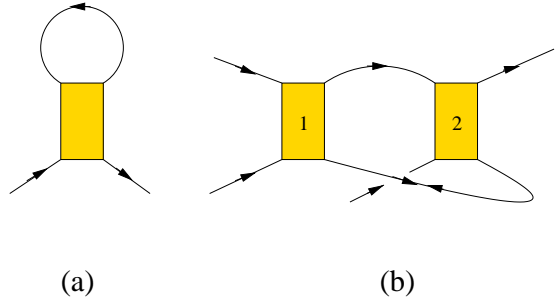,width=0.6\textwidth}
\end{center}
\caption{Examples of contributions that vanish in the thermodynamic limit
of the reduced BCS model. (a) the lowest order normal selfenergy.
(b) A second order term with three fermions going in and one 
coming out.}
\label{appfig1}
\end{figure}

The first example is the contribution to the right hand side
of the flow equation for the selfenergy, eq.\ \Ref{eq:3Si},
from the normal Fock graph shown in Figure \ref{appfig1} (a). 
This is a term that changes the normal part of the selfenergy,
i.e.\ it shifts the Fermi surface and changes the quasiparticle weight.
Its value is obtained by applying the part of the Laplacian to the vertex,
\begin{equation}
\cT =
- \left(
\frac{\de}{\de \psq}\; , \; 
\bar \lQ b_S \frac{\de}{\de \ps} 
\right)
\;
\frac{v}{\Om} \, \bar X \, X .
\end{equation}
Straightforward evaluation of the derivatives gives
\begin{equation}
\cT 
=
\frac{v}{\Om}\;
\frac{1}{\Om}
\sum_{k,\si}
(\bar\lQ b_S)_{\si,\si} (-k)\; 
\psq_\si (k) \ps_\si (k) 
\end{equation}
which is of order $\Om^{-1}$ as compared to the term in the action,
hence vanishes for $\Om \to \infty$. 

The second example is the contribution to the right hand side of the
flow equation \Ref{eq:3Four} for the anomalous part of the vertex 
where three particles go in and one comes out, shown in 
Figure \ref{appfig1} (b), with value
\begin{equation}
\left(\frac{v}{\Om}\right)^2\;
\BEqual{
\sfrac{\del}{\del s}
\left\lbrack
\left(
\sfrac{\de}{\de \psq^{(2)}}\, ,\;
\bar \lQ\, b_{G}
\sfrac{\de}{\de \ps^{(1)}}
\right)\;
\left(
\sfrac{\de}{\de \ps^{(1)}}\, ,\;
\veps\, a_{G}
\sfrac{\de}{\de \ps^{(2)}}
\right)\;
\right\rbrack
\;
\bar X ^{(1)}
X ^{(1)}
\bar X ^{(2)}
X ^{(2)}
}
\end{equation}
Straightforward evaluation of the action of the Laplacian gives
\begin{equation}\label{eq:subfour}
\left(
\frac{v}{\Om}
\right)^2
\bar X ^{(1)}\; Y
=
\frac{1}{\Om}\;
\frac{v^2}{\Om}
\bar X ^{(1)}\; Y
\end{equation}
with 
\begin{equation}
Y
=
\frac{1}{\Om}\;
\sum_{k}
f(\vec k)^3\;
\left\lbrack
\frac{\del}{\del s} (b_G(k) a_G(k))
\right\rbrack
(\I \om - e(\vec k))
\sum_{\si}
\psq_\si^{(2)} (k) \, \psq_\si^{(2)} (k) .
\end{equation}
Like $X$ and $\bar X$, $Y$ is a bilinear in the fermion fields with 
a similar momentum structure. However, the prefactor
in \Ref{eq:subfour} contains an 
additional factor $1/\Om$ as compared to the main term. 

In both examples, the extra power $1/\Omega$ arises because
there is no summation over loop momenta -- the structure of the 
graphs and the vertices of the reduced BCS model fix the loop momenta
in terms of the external ones. Then there remains an overall $1/\Om$ 
which makes the terms subleading in $\Om$. 

\subsection{The general argument}
It is now easy to generalize this argument to general graphs. 
The Cooper pair terms $X$ and $\bar X$ can be rewritten as 
\begin{eqnarray}
X 
&=&
\frac{1}{\Om^2}
\sum_{k,l} f(\vec k) \;
\ps_+ (k) \ps_- (l) \de_\Om(k,-l)
\\
\bar X 
&=&
\frac{1}{\Om^2}
\sum_{k',l'} f(\vec k') \;
\psq_- (k') \psq_+ (l') \de_\Om(k',-l')
\end{eqnarray}
which makes it clear that quantities of order $1$ contain 
one sum over momenta and frequencies per field $\ps$ 
or $\psq$, if all momentum constraints are written in terms of 
the delta functions $\de_\Om$. By definition of $\de_\Om$, 
\begin{equation}\label{eq:deOmsq}
\left(\de_\Om (k,k')\right)^2
=
\Om \de_\Om (k,k')
\end{equation}
Because the initial vertex function is $O(\Om^{-1})$, 
a term of order $p$ in the vertex is of order $\Om^{-p+\ell}$ 
where $\ell$ is the number of times two delta functions get paired
so as to give a square as in \Ref{eq:deOmsq}.
To get a contribution of order $\Om^{-1}$ to 
the two--particle vertex, $\ell$ must equal $p-1$. 
Such a complete pairing occurs only in ladder graphs. 
Similarly, the $O(1)$ contribution to the selfenergy 
can come only from anomalous Fock--type diagrams for the 
selfenergy made by joining two external legs of a ladder graph. 

The argument we just gave does not require that the model is doubly reduced. 
All these graphical arguments apply directly to the reduced BCS 
model in Hamiltonian form, i.e.\ with full frequency dependence.
It should be noted that the diagrammatic approach, i.e.\
determining for each diagram separately whether its contribution
is $O(1)$ or subleading, is not a proof in the mathematical sense
that this gives the exact solution
because a graph--by--graph analysis presupposes 
that the perturbative series is convergent
or at least asymptotic (the functional integral argument in
\cite{Muehlschlegel} also involves an unjustified exchange of the thermodynamic 
limit with the limit of continuous Euclidian time). 
A complete mathematical proof that this gives
the solution of the reduced BCS model in Hamiltonian form
was given in \cite{Thirring}; see also \cite{Hemmen}.

\end{appendix}

\bigskip\noindent
{\bf Acknowledgement. } 
We would like to thank Tilmann Enss and Andrei Katanin for discussions,
Walter Wreszinski for bringing \cite{Thirring} and \cite{Hemmen} to 
our attention, 
and the Erwin--Schr\" odinger Institute for its hospitality during the 
completion of this work.

\end{document}